# Arbitrage in Fractal Modulated Black-Scholes Models When the Volatility is Stochastic

Erhan Bayraktar [*]     H. Vincent Poor [†]

## Abstract

In this paper an arbitrage strategy is constructed for the modified Black-Scholes model driven by fractional Brownian motion or by a time changed fractional Brownian motion, when the volatility is stochastic. This latter property allows the heavy tailedness of the log returns of the stock prices to be also accounted for in addition to the long range dependence introduced by the fractional Brownian motion. Work has been done previously on this problem for the case with constant 'volatility' and without a time change; here these results are extended to the case of stochastic volatility models when the modulator is fractional Brownian motion or a time change of it. (Volatility in fractional Black-Scholes models does not carry the same meaning as in the classic Black-Scholes framework, which is made clear in the text.)

Since fractional Brownian motion is not a semi-martingale, the Black-Scholes differential equation is not well-defined sense for arbitrary predictable volatility processes. However, it is shown here that any almost surely continuous and adapted process having zero quadratic variation can act as an integrator over functions of the integrator and over the family of continuous adapted semi-martingales. Moreover it is shown that the integral also has zero quadratic variation, and therefore that the integral itself can be an integrator. This property of the integral is crucial in developing the arbitrage strategy. Since fractional Brownian motion and a time change of fractional Brownian motion have zero quadratic variation, these results are applicable to these cases in particular. The appropriateness of fractional Brownian motion as a means of modeling stock price returns is discussed as well.

[*]Department of Mathematics, University of Michigan, 2074 East Hall, Ann Arbor, MI 48109-1109, erhan@umich.edu

[†]Department of Electrical Engineering, Princeton University, Princeton, NJ 08544, poor@princeton.edu

# 1 INTRODUCTION

The classic Black-Scholes (B-S) model for market risk does not account for long-range dependence in asset prices, although there is strong empirical evidence which shows that stock market price dynamics possess this property (e.g. Bayraktar et al. (2004), Bayraktar, Poor and Sircar (2003), Cutland et al. (1995), Greene and Fielitz (1977), Teverovsky et al. (1999), and Willinger et al. (1999).) Therefore, there have been efforts in recent years to model stock market prices using a modified B-S model where the modulating process is not Brownian motion, but is rather a fractional Brownian motion (fBm). See for example Cutland et al. (1995), Salopek (1998), Maheswaran and Sims (1992) and Hu and Øksendal (2003). Since fBm is not a semi-martingale, see e.g. Rogers (1997), there is no equivalent martingale measure for fBm, and therefore in a frictionless market where continuous trading is possible there exist arbitrage strategies.

And indeed, for fBm-modulated Bachelier models, Rogers (1997), and for fBm-modulated B-S models with constant volatility, Cheridito (2003) and Shiryaev (1998), have constructed explicit arbitrage strategies.

There are at least two possible frameworks that one can adopt for giving meaning to fractional B-S models with stochastic volatility. (In what follows we will make it clear what is meant by volatility.) These frameworks differ in the way they define integrals. In the framework we choose, the stochastic integrals are understood as the probabilistic limits of Stieltjes sums. That is, given stochastic processes $Y$ and $X$, such that $Y$ is adapted to the filtration generated by $X$, we say that the integral $\int Y \, dX$ exists if for any $t < \infty$, and for each sequence of partitions $\{\sigma^n\}_{n \in \mathbb{N}}$, $\sigma^n = (T_1^n, T_2^n, ..., T_{k_n}^n)$, of the interval $[0, t]$ that satisfies $\lim_{n \to \infty}, \max_i |T_{i+1}^n - T_i^n| = 0$, the sequence of sums $\left( \sum_i Y_{T_i^n} (X_{T_{i+1}^n} - X_{T_i^n}) \right)$

---

[1]**Key Words:** Fractional Brownian Motion, Arbitrage, Stochastic Volatility, Stochastic Integration, Fractal Market Models

[2]JEL Classification: G19

[3]2000 Mathematics Subject Classification. 62P05 60G15, 60G18, 60H05

converges in probability. That is, we define

$$\int_0^t Y_s dX_s = \mathbb{P} - \lim_{n \to \infty} \sum_i Y_{T_i^n} (X_{T_{i+1}^n} - X_{T_i^n}). \tag{1.1}$$

Hu and Øksendal (2003) have considered another framework for the fractional B-S model with the difference that the integrals in their framework are Wick type integrals. The Wick type integral of a process $Y$ with respect to a process $X$ is defined as

$$\int_0^t Y_{T_i^n} \diamond (X_{T_{i+1}^n} - X_{T_i^n}) \tag{1.2}$$

where the convergence is convergence in the $L^2$ space of random variables. (The Wick product is defined using the tensor product structure of $L^2$; see Bayraktar and Poor (2002).) The Wick type integral of $Y$ with respect to fBm $X = B^H$, $H \in [1/2, 1)$ (see below for a definition of fractional Brownian motion) is equal to Stieltjes integral plus a drift term (see Duncan et al. (2000), Thm. 3.12),

$$\int_0^t Y_s dB_s^H = \int_0^t Y_s \diamond dB_s^H + \int_0^t D_s^\phi Y_s ds, \tag{1.3}$$

where $\phi(s,t) = H(2H-1)|s-t|^{2H-2}$, and $D_s^\phi Y_t := (D^\phi Y_t)(s)$ is the Hida derivative of the random variable $Y_t$.

The Wick integral defined by (1.2), has zero mean. Consequently, the stochastic differential term of a stochastic differential equation that is written in terms of wick integrals does not contribute to the mean rate of change. Thence in some cases modeling the dynamics of the states of a phenomenon with Wick-type differential equations makes sense, and one can then employ the powerful tools of fractional noise calculus. (See for example Bayraktar and Poor (2002), Brody et al. (2000).) Application of the fractional noise calculus to stock price modeling must be done with some caution, however, since in finance, one must make economic sense of the definition of gain associated with a trading strategy. Hu and Øksendal (2003) claim that the fractional B-S model they consider does not lead to arbitrage opportunities. This claim is, however, based on the redefinition of the class of self-financing strategies. The non-deterministic self-financing strategies in a Stieltjes framework are no longer self-financing strategies in a Wick framework as a simple application of (1.3) shows, so all of the self-financing arbitrage strategies of the Stieltjes framework are ruled out by the approach of Hu and Øksendal (2003). However in the Wick framework it is difficult to give economic interpretations to gain processes associated with trading strategies. To see this consider the following discrete strategy with the initial holding being x dollars. At time $t = 0$, the strategy requires one to enter the market. However no trade is made. At time $t = 1$, the strategy tells one to buy stock at price $S_1$ with all the available capital, and keep this portfolio until time $t = 2$, given that the stock price is $S_2$ at that time. Now one can ask: what the gain associated with this strategy? The answer to this question should be independent of the underlying model for the prices and should be $(x/S_1)(S_2 - S_1)$. However in the fractional B-S model of Øksendal and Hu the gain prescribed to the portfolio above is $(x/S_1)(S_1 \diamond S_2)$. It is difficult to attach a clear economic meaning to this quantity since the Wick product is not a path-wise product but rather is defined using the tensor product structure of the $L^2$ space of random variables. Note also that the gain $(x/S_1)(S_2 - S_1)$ is an outcome of a trading strategy which is self-financing in the sense that the current value of the portfolio is equal to the initial holdings plus the trading gains. However no self-financing strategy in the sense of Øksendal and Hu is able to reproduce the gain $(x/S_1)(S_2 - S_1)$, which can easily be seen by the application of (1.3) and using the fact that the trading strategy is not deterministic. Hence the no-arbitrage conclusion of Hu and Øksendal (2003), which arises as a result of redefining self-financing strategies, cannot be interpreted within the usual meaning of this term, and thus we prefer to apply the definition (1.1). (See Björk and Hult (2003) and Sottinen and Valkeila (2003), who also argue that Wick type integrals are not suitable for defining trading strategies.)

The previous results of Rogers, Cheridito, and Shiryaev on fBm-modulated B-S markets have considered only the situation in which the 'volatility' is constant. It is natural to ask whether the presence of stochastic volatility or time change might remove the possibility of arbitrage. Stochastic volatility can also be used to model heavier tails in the returns distributions, along with the existing long range dependence. Heavy tailed marginals for stock price returns have been observed in many empirical studies since the early 1960's (e.g., Greene and Fielitz (1977), Mandelbrot (1963)). We will seek an answer to the question of whether arbitrage arises in a frictionless market when both the heavy tailedness of the marginals of the stock price returns and the long range dependence of the same series are accounted for, by considering a fractal version of the standard Black-Scholes model with stochastic volatility, and we will generalize the result of Shiryaev (1998) to this situation. The explicit construction of arbitrage strategies will reveal how the arbitrage arises in these models, which will pave the way for engineering models that can explain the long range dependence observed in the stock price indices without giving rise to arbitrage.

We will analyze two cases. In the first case the volatility process is a function of the modulating process, and in the second case it is an Itô process satisfying a stochastic differential equation driven by a Brownian motion. The most common examples of this latter type are the log-normal stochastic volatility model of Hull and White (1987), the mean-reverting model of Stein and Stein (1991), and the Cox-Ingersoll-Ross (CIR) model of Heston (1993).

To study this problem, we consider a market with one risky security whose price evolves according to a stochastic process $(Y_t)_{t \in [0,T]}$, and a riskless asset that evolves according to the deterministic process $(X_t)_{t \in [0,T]}$. We will assume the following model for the market:

$$
\begin{aligned}
X_t &= \exp(rt) \qquad , \ 0 \le t \le T \\
dY_t &= Y_t \left( \nu dt + \sigma_t Z_t \right) \qquad , \ 0 \le t \le T
\end{aligned}
\tag{1.4}
$$

where $\sigma$ is a stochastic volatility process, and $Z$ denotes a process with zero quadratic variation. We are interested in the case when $Z_t = B_t^H$ or $Z_t = B_{A_t}^H$ where $B^H$ is a fractional Brownian motion (fBm) and $A$ is a continuous non-decreasing process. FBm, $B^H$ is a Gaussian random process with $E\left\{ B_t^H \right\} = 0$ and $E\left\{ B_t^H B_s^H \right\} = \frac{1}{2} \left( |t|^{2H} + |s|^{2H} - |t-s|^{2H} \right)$ where $H \in (0,1]$ is the so-called Hurst parameter. (Note that $H = \frac{1}{2}$ gives standard

Brownian motion, and that for $H \notin (0, 1]$, the auto-correlation function is not semi-definite.) FBm models are able to capture long range dependence in a parsimonious way. Consider for example the fractional Gaussian noise $N(k) := B^H(k) - B^H(k - 1)$. The auto-correlation function of $N$, which is denoted by $r$, satisfies the asymptotic relation

$$r(k) \sim r(0)H(2H - 1)k^{2H-2}, \quad \text{as } k \to \infty. \tag{1.5}$$

For $H \in (1/2, 1]$, $Z$ exhibits long-range dependence, which is also called the Joseph effect in the terminology of Mandelbrot (1997). For $H = 1/2$ all correlations at non-zero lags are zero. For $H \in (0, 1/2)$ the correlations are summable, and in fact they sum to zero. The latter case is less interesting for financial applications, since the empirical evidence suggests positive correlation for the log return series. Therefore we focus on the case *persistent* fBm case, i.e. the case in which $H \in (1/2, 1]$.

The second equation of (1.4) should not be interpreted in the same way as an integral with respect to Brownian motion. Since the stochastic integral $\int_0^t \sigma_s dB_s^H$ is not a martingale, it has a non-zero mean; hence the predictable component of the log price process is not contained only in the drift term. Moreover $\sigma$ is not the quadratic variation of the log price process. Therefore $\sigma$ in (1.4) is not to be interpreted as a volatility in the usual meaning of this term. Rather than having a meaning as volatility, it serves as a means of incorporating heavy tailed marginals for the log return distribution. However in this text we will still refer to $\sigma$ as the volatility process for convenience.

The integrals of predictable processes with respect to fBm may not converge in probability, which is an immediate consequence of the Bichteler-Dellacherie Theorem (see e.g. Protter (1990)). [1] But, by using an integration-by-parts argument we will find a class of processes that can be integrated with respect to fBm, and which is large enough

---

[1] Basically, this theorem states that a good integrator must be a semi-martingale, i.e. for an integral operator to be well defined for the class of adapted processes the integrator must be a semi-martingale (see Protter (1990).)) However fBm is not a semi-martingale as a consequence of the Burkholder-Davis inequality since its quadratic variation is infinite for $H \in (0, 1/2)$ and is zero for $H \in (1/2, 1]$.

for our purposes.

The main idea of this paper is to develop arbitrage strategies for the market modeled by (1.4) whenever this model is well-defined; i.e. for all the volatility processes such that (1.4) can be given a meaning. As we will see, the family of volatilities for which (1.4) is meaningful is large enough to include the stochastic volatility models in the literature, e.g. the stochastic volatility models of Fouque et al. (2000), Heston (1993), Hull and White (1987), Stein and Stein (1991). Our treatment develops in such a way that the results are true not only with fractional Brownian motion and time change of it as the modulating process but also for any modulator that is almost surely continuous and whose associated quadratic variation process is a.s. zero.

It should be noted that the existence of arbitrage opportunities for a fractional B-S model does not rule out the use of these models as candidates for stock price modeling, as the arbitrage strategies exist only in frictionless markets. For the sake of coherence, we defer the discussion of this issue until the concluding section.

We present our results in two main sections. The first, Section 2, deals with constructing arbitrage strategies in the market modeled by (1.4) under some restrictions on the volatility process. The second, Section 3, presents two different families of volatility processes of interest satisfying the assumptions needed in Section 2. Finally, in Section 4, we discuss our results briefly and offer some conclusions.

## 2   EXISTENCE OF ARBITRAGE

Throughout this treatment we assume that we are given a filtered, complete probability space $(\Omega, \mathcal{F}, (\mathcal{F}_t)_{t \in [0,T]}, P)$ satisfying the usual hypothesis:

**(i)** $\mathcal{F}_0$ contains all the $P$-null sets of $\mathcal{F}$, and

**(ii)** $\mathcal{F}_t = \bigcap_{u > t} \mathcal{F}_u$ for all $t \in [0, t]$(i.e. the filtration $(\mathcal{F}_t)_{t \in [0,T]}$ is right-continuous).

For any a.s. continuous process $(Z_t)_{t \in [0,T]}$, the associated quadratic variation process is defined as

$$[Z,Z]_t = \text{P-} \lim_{|\Delta| \to 0} \sum_{i=0}^{n-1} (Z_{t_{i+1}} - Z_{t_i})^2,$$

where $\Delta = \{(t_0, t_1), (t_1, t_2), ..., (t_{n-1}, t_n)\}$ denotes any partition of $[0,t]$, and where P-lim denotes the limit in probability.

We begin with a basic lemma, which will be an important tool in constructing an arbitrage strategy. [2]

**Lemma 2.1** (A Modified Itô Formula): *Suppose that the function* $F : [0,T] \times \mathbb{R} \longrightarrow \mathbb{R}$ *has continuous partial derivatives of order 2 and* $(Z_t)_{t \in [0,T]}$ *is an a.s. continuous process with zero quadratic variation. Then the following modified Itô formula holds:*

$$dF(t, Z_t) = \partial_1 F(t, Z_t) dt + \partial_2 F(t, Z_t) dZ_t. \tag{2.6}$$

*Proof:* First let us write, for fixed $t \in [0,T]$, and for an arbitrary partition $\Delta$ of $[0,t]$,

$$F(t, Z_t) = F(t_0, Z_{t_0}) + \sum_{i=0}^{n-1} F(t_{i+1}, Z_{t_{i+1}}) - F(t_i, Z_{t_i}).$$

---

[2] Our formulation differs from that of Ruzmaikina (1999) and Zähle (1998) since these works rely on the fact that the integrator is Hölder continuous of order $H \in (1/2, 1]$. Among the functions with compact support the set of functions having finite $1/H$ variation is a super set of the set of functions with Hölder continuity exponent $H$, since any bounded $1/H$ variation function can be obtained by a bounded non-decreasing time change of a function with Hölder exponent $H$. Hence our results hold for a larger set of integrators; for example we can consider a time changed fractional Brownian motion as an integrator. Also note that these works consider almost sure convergence, while we consider convergence in probability. As we shall see, this difference allows us to make sense of the integrals of adapted semi-martingales in our framework. This property is essential since all the volatility models introduced in the literature are semi-martingales.

Applying Taylor's formula to each summand we have

$$
\begin{aligned}
F(t, Z_t) = {} & F(t_0, Z_{t_0}) + \sum_{i=0}^{n-1} \Bigg[ \partial_1 F(t_i, Z_{t_i})(t_{i+1} - t_i) \\
& + \partial_2 F(t_i, Z_{t_i})(Z_{t_{i+1}} - Z_{t_i}) + 1/2 \Big\{ \partial_{11} F(t_i^*, \xi_i)(Z_{t_{i+1}} - Z_{t_i})^2 \\
& + 2\partial_{12} F(t_i^*, \xi_i)(Z_{t_{i+1}} - Z_{t_i})(t_{i+1} - t_i) + \partial_{22} F(t_i^*, \xi_i)(t_{i+1} - t_i)^2 \Big\} \Bigg],
\end{aligned}
$$

where $\xi_i$ is between $Z_{t_i}$ and $Z_{t_{i+1}}$, and $t_i^*$ is between $t_i$ and $t_{i+1}$. Note that

$$
\left| \sum_{i=0}^{n-1} \partial_{11} F(t_i^*, \xi_i)(Z_{t_{i+1}} - Z_{t_i})^2 \right| \;\leq\; M_1 \sum_{i=0}^{n-1} (Z_{t_{i+1}} - Z_{t_i})^2, \tag{2.7}
$$

$$
\left| \sum_{i=0}^{n-1} \partial_{12} F(t_i^*, \xi_i)(Z_{t_{i+1}} - Z_{t_i})(t_{i+1} - t_i) \right| \leq M_2 \sqrt{\sum_{i=0}^{n-1} (Z_{t_{i+1}} - Z_{t_i})^2 \sum_{i=0}^{n-1} (t_{i+1} - t_i)^2}, \tag{2.8}
$$

and

$$
\left| \sum_{i=0}^{n-1} \partial_{22} F(t_i^*, \xi_i)(t_{i+1} - t_i)^2 \right| \leq M_3 \sum_{i=0}^{n-1} (t_{i+1} - t_i)^2, \tag{2.9}
$$

where

$$
\begin{aligned}
M_1 &= \sup\{| \partial_{11} F(p, Z_s) |: 0 \leq s \leq t, 0 \leq p \leq t\}, \\
M_2 &= \sup\{| \partial_{12} F(p, Z_s) |: 0 \leq s \leq t, 0 \leq p \leq t\}, \\
M_3 &= \sup\{| \partial_{22} F(p, Z_s) |: 0 \leq s \leq t, 0 \leq p \leq t\}.
\end{aligned}
$$

Since $Z$ and the second partials of $F$ are continuous, $M_1, M_2,$ and $M_3$ are a.s. finite random variables.

We will show that (2.7), (2.8) and (2.9) converge in probability (i.p.) to zero as $| \Delta | \longrightarrow 0$. By our assumption $Z$ has zero quadratic variation. So the terms multiplying $M_1$, $M_2$ and $M_3$ in (2.7), (2.8), and (2.9), respectively, converge i.p. to zero. The following lemma will be sufficient to conclude that the expressions of (2.7), (2.8), and (2.9) also converge i.p. to zero.

**Lemma 2.2** *Suppose $(Z_n)$ is a sequence of positive random variables converging to zero in probability, and $M$ is an a.s finite and positive random variable. Then $(MZ_n)$ converges in probability to zero.*

*Proof:* The proof is straightforward, but we include it for the sake of completeness. For any $\epsilon > 0$, we can write

$$
\begin{aligned}
P(Z_n M \geq \epsilon) &= \sum_j P(Z_n M \geq \epsilon, j - 1 \leq M < j) \qquad (2.10) \\
&\leq \sum_j P(Z_n j \geq \epsilon, j - 1 \leq M < j).
\end{aligned}
$$

We wish to take the limit as $n \longrightarrow \infty$ of the left-hand side of (2.6). Since

$$
P(Z_n j \geq \epsilon, j - 1 \leq M < j) \leq P(j - 1 \leq M < j),
$$

the dominated convergence theorem allows us to interchange the limit and summation. Then, since $(Z_n)$ converges in probability to 0, we have

$$
\lim_{n \to \infty} \sum_j P(Z_n j \geq \epsilon, j - 1 \leq M < j) = \sum_j \lim_{n \to \infty} P(Z_n j \geq \epsilon, j - 1 \leq M < j) = 0,
$$

and Lemma 2.2 follows.                                                                    □

Now the only thing that remains to be proven for (2.6) is the convergence in probability of $\sum_{i=0}^{n-1} \partial_2 F(t_i, Z_{t_i})(Z_{t_{i+1}} - Z_{t_i})$. This convergence is due to Lin (1995), from which it follows that

$$
\lim_{\Delta \to 0} \sum_{i=0}^{n-1} \partial_2 F(t_i, Z_{t_i})(Z_{t_{i+1}} - Z_{t_i}) = \int_0^{Z_t} \partial_2 F(t, x)dx - \int_0^t \int_0^{Z_s} \partial_{22} F(x, s)dxds.
$$

This completes the proof of the main lemma.                                              □

We will now specify an arbitrage strategy for the market modeled by (1.4) with an additional assumption that will allow us to use the modified Itô formula in this model. The main result is summarized in the following.

**Theorem 2.1** *(An Arbitrage Portfolio): Consider a market modeled by (1.4). Let Z be an almost surely continuous process with zero quadratic variation. We will assume that either the filtration is generated by Z or it is finer. Suppose $(\sigma_t)$ is an adapted process such that $(\int_0^t \sigma_s dZ_s)_{t \in [0,T]}$ exists, has zero quadratic variation, and is continuous. Then, for each $c > 0$, the following portfolio is an arbitrage strategy*

$$\vartheta_t^0 = \frac{c}{Y_0}\left(Y_0^2 - \left(e^{-rt}Y_t\right)^2\right), \quad \vartheta_t^1 = \frac{2c}{Y_0}\left(Y_t e^{-rt} - Y_0\right). \tag{2.11}$$

*where the quantities $\vartheta_t^0$, and $\vartheta_t^1$ are the number of units of riskless asset and of the risky asset, respectively, held at time t.*

*Remark*: Note that $Z_t = B_t^H$, or $Z_t = B_{A_t}^H$ for a process $A$ that is continuous and non-decreasing and continuous, are two special cases for which this theorem can be applied. From (2.11) it can be seen that to implement the arbitrage portfolio one does not need to estimate $H$, the portfolio depends on $H$ only through the price.

*Proof:* First notice that the modified Itô formula of (2.6) implies that we have $Y_t = Y_0 \exp\left(\nu t + \int_0^t \sigma_s dZ_s\right)$, and thus (2.11) can be written as

$$\begin{aligned}
\vartheta_t^0 &= cY_0\left(1 - \exp\left(2(\nu - r)t + 2\int_0^t \sigma_s dZ_s\right)\right) \\
\vartheta_t^1 &= 2c\left(\exp\left((\nu - r)t + \int_0^t \sigma_s dZ_s\right) - 1\right),
\end{aligned} \tag{2.12}$$

Let us denote by $(P_t)_{t \in [0,T]}$ the value process of the portfolio (2.11). We will first show that $P_t > 0, \forall t$, and then we will proceed by showing that the given portfolio is

self-financing. We have

$$
\begin{aligned}
P_t &= \vartheta_t^0 \exp(rt) + \vartheta_t^1 Y_t \\
&= cY_0 \left( 1 - \exp\left( 2(\nu - r)t + 2\int_0^t \sigma_s dZ_s \right) \right) \exp(rt) \\
&\quad + 2c \left( \exp\left( (\nu - r)t + \int_0^t \sigma_s dZ_s \right) - 1 \right) Y_0 \exp\left( \nu t + \int_0^t \sigma_s dZ_s \right) \\
&= \left( cY_0 \exp(rt) \left( 1 - \exp\left( 2(\nu - r)t + 2\int_0^t \sigma_s dZ_s \right) \right) \right. \\
&\quad + 2\exp\left( 2(\nu - r)t + 2\int_0^t \sigma_s dZ_s \right) \\
&\quad \left. - 2\exp\left( (\nu - r)t + \int_0^t \sigma_s dZ_s \right) \right) \\
&= cY_0 \exp(rt) \left( \exp\left( (\nu - r)t + \int_0^t \sigma_s dZ_s \right) - 1 \right)^2 > 0,
\end{aligned}
\tag{2.13}
$$

which is the first step. We will now show that the portfolio is self financing. By (2.13) we see that $P_t$ is a smooth enough function of $t$ and $\int_0^t \sigma_s dZ_s$, both of which have zero quadratic variation and are continuous. Therefore using the modified Itô formula we have

$$
\begin{aligned}
dP_t &= crY_0 \exp(rt) \Bigg( \left( \exp\left( (\nu - r)t + \int_0^t \sigma_s dZ_s \right) - 1 \right)^2 \\
&\quad + 2c(\nu - r)Y_0 \exp(rt) \left( \exp\left( (\nu - r)t + \int_0^t \sigma_s dZ_s \right) - 1 \right) \\
&\quad \times \exp\left( (\nu - r)t + \int_0^t \sigma_s dZ_s \right) \Bigg) dt \\
&\quad + 2cY_0 \exp(rt) \left( \exp\left( (\nu - r)t + \int_0^t \sigma_s dZ_s \right) - 1 \right) \\
&\quad \times \exp\left( (\nu - r)t + \int_0^t \sigma_s dZ_s \right) \sigma_t dZ_t.
\end{aligned}
\tag{2.14}
$$

On the other hand we have

$$
\vartheta_t^0 dX_t + \vartheta_t^1 dY_t = \vartheta_t^0 r \exp(rt)dt + \vartheta_t^1 Y_t \left( \nu dt + \sigma_t dZ_t \right)
\tag{2.15}
$$

$$= \left( \vartheta_t^0 r \exp(rt) + \vartheta_t^1 Y_t \nu \right) dt + \vartheta_t^1 Y_t \sigma_t dZ_t$$

$$= \left\{ cY_0 \left( 1 - \exp\left( 2(\nu - r)t + 2\int_0^t \sigma_s dZ_s \right) \right) r \exp(rt) \right.$$
$$\left. + 2c\nu \left( \exp\left( (\nu - r)t + \int_0^t \sigma_s dZ_s \right) - 1 \right) Y_0 \exp\left( \nu t + \int_0^t \sigma_s dZ_s \right) \right\} dt$$
$$+ \left\{ 2c \left( \exp\left( (\nu - r)t + \int_0^t \sigma_s dZ_s \right) - 1 \right) Y_0 \exp\left( \nu t + \int_0^t \sigma_s dZ_s \right) \right\} \sigma_t dZ_t$$

$$= \left\{ cY_0 r \exp(rt) \left( 1 - \exp\left( 2(\nu - r)t + 2\int_0^t \sigma_s dZ_s \right) \right) \right.$$
$$+ 2cY_0(\nu - r) \exp(rt) \left( \exp\left( (\nu - r)t + \int_0^t \sigma_s dZ_s \right) - 1 \right) \exp\left( (\nu - r)t + \int_0^t \sigma_s dZ_s \right)$$
$$\left. + 2cY_0 r \exp(rt) \left( \exp\left( (\nu - r)t + \int_0^t \sigma_s dZ_s \right) - 1 \right) \exp\left( (\nu - r)t + \int_0^t \sigma_s dZ_s \right) \right\} dt$$
$$+ 2cY_0 \exp(rt) \left( \exp\left( (\nu - r)t + \int_0^t \sigma_s dZ_s \right) - 1 \right)$$
$$\times \exp\left( (\nu - r)t + \int_0^t \sigma_s dZ_s \right) \sigma_t dZ_t$$

$$= \left\{ cY_0 r \exp(rt) \left( \left( 1 - \exp\left( 2(\nu - r)t + 2\int_0^t \sigma_s dZ_s \right) \right) \right. \right.$$
$$\left. + 2 \left( \exp\left( 2(\nu - r)t + 2\int_0^t \sigma_s dZ_s \right) \right) - 2\exp\left( (\nu - r)t + \int_0^t \sigma_s dZ_s \right) \right)$$
$$\left. + 2cY_0(\nu - r) \exp(rt) \left( \exp\left( (\nu - r)t + \int_0^t \sigma_s dZ_s \right) - 1 \right) \exp\left( (\nu - r)t + \int_0^t \sigma_s dZ_s \right) \right\} dt$$
$$+ 2cY_0 \exp(rt) \left( \exp\left( (\nu - r)t + \int_0^t \sigma_s dZ_s \right) - 1 \right)$$
$$\times \exp\left( (\nu - r)t + \int_0^t \sigma_s dZ_s \right) \sigma_t dZ_t.$$

Since the right-hand side of (2.15) is equal to the right-hand side of (2.14), the portfolio we have constructed is self-financing. Having a self-financing strategy that is at all times positive means that we have an arbitrage strategy. □

# 3   EXISTENCE OF VOLATILITY PROCESSES SATISFYING THE ASSUMPTIONS OF THEOREM 2.1

In this section we will see that the assumptions of Theorem 2.1 are not void, and that there is actually a large class of processes satisfying the required assumptions. We consider two separate types of stochastic volatility models: those in which the volatility is a semi-martingale, and those in which it is a function of the asset-modulating fBm.

## 3.1   WHEN THE VOLATILITY PROCESS IS A SEMI-MARTINGALE

To consider the case of semi-martingale volatility, we begin with the following result that will lead us to integration by parts.

**Lemma 3.1** *Suppose* $\Phi : \mathbb{R}^2 \to \mathbb{R}$ *has continuous partial derivatives of order 2, $W$ is an adapted a.s. continuous semi-martingale, and $Z$ is an adapted a.s. continuous process with zero quadratic variation. Then $\int_0^t \Phi'_x dZ_s$ can be defined via the following relationship*

$$
\begin{aligned}
\Phi(Z_t, W_t) - \Phi(Z_0, W_0) \;=\; & \int_0^t \Phi'_x(Z_s, W_s) dZ_s + \int_0^t \Phi'_y(Z_s, W_s) dW_s \\
& + \frac{1}{2} \int_0^t \Phi''_{yy}(Z_s, W_s) d[W, W]_s \;,
\end{aligned}
$$

*and therefore it is a.s. continuous.*

*Proof:* Let $\Delta = \{(t_0, t_1), (t_1, t_2), ..., (t_{n-1}, t_n)\}$ be any partition of $[0, t]$, and write

$$
\Phi(Z_t, W_t) = \sum_{(t_{i-1}, t_i)} [\Phi(Z_{t_i}, W_{t_i}) - \Phi(Z_{t_{i-1}}, W_{t_{i-1}})].
$$

Applying Taylor's formula to each summand we have

$$
\begin{aligned}
\Phi(Z_t, W_t) \;=\; & \sum_{(t_{i-1}, t_i)} \Phi_x(Z_{t_{i-1}}, W_{t_{i-1}})(Z_{t_i} - Z_{t_{i-1}}) \\
& + \sum_{(t_{i-1}, t_i)} \Phi_y(Z_{t_{i-1}}, W_{t_{i-1}})(W_{t_i} - W_{t_{i-1}}) \\
& + \frac{1}{2} \sum_{(t_{i-1}, t_i)} \Phi_{xx}(\xi_i, \eta_i)(Z_{t_i} - Z_{t_{i-1}})^2 \\
& + \sum_{(t_{i-1}, t_i)} \Phi_{xy}(\xi_i, \eta_i)(Z_{t_i} - Z_{t_{i-1}})(W_{t_i} - W_{t_{i-1}}) \\
& + \frac{1}{2} \sum_{(t_{i-1}, t_i)} \Phi_{yy}(\xi_i, \eta_i)(W_{t_i} - W_{t_{i-1}})^2.
\end{aligned}
$$

Since $\Phi_{xx}, \Phi_{xy}, \Phi_{xy}, Z$ and $W$ are continuous, Lemma 3.1 follows from the same type of arguments as those used in the proof of Lemma 2.1. $\qquad\square$

Lemma 3.1 assures that integration by parts, which will help us define integrals of semi-martingales with respect to continuous processes of zero quadratic variation, holds. In particular, we can state the following.

**Corollary 3.1** (INTEGRATION-BY-PARTS)*: Suppose $W$ and $Z$ are as in Lemma 3.1. Then $(\int_0^t W_s dZ_s)$ exists as a limit in probability of sums over finite partitions, is a.s. continuous, and is given by*

$$
\int_0^t W_s dZ_s = Z_t W_t - \int_0^t Z_s dW_s.
$$

*Proof:* Using Lemma 3.1 we have an immediate conclusion by choosing $\Phi(W_t, Z_t) = W_t Z_t$. $\square$

Having found the existence of the integral of a continuous semi-martingale with respect to an a.s. continuous process with zero quadratic variation, we will proceed to show that the integral itself has zero quadratic variation, which is one of the requirements of Theorem 2.1. This property of the integral also implies that the integral of an a.s. continuous and adapted semi-martingale with respect to the integral process is well defined, meaning that we can perform repeated integrations.

**Lemma 3.2** : *Suppose $W$ is an a.s. continuous and adapted semi-martingale, and $Z$ is an a.s. continuous process having zero quadratic variation. Then $(\int_0^t W_s dZ_s)$ has zero quadratic variation.*

*Proof:* The proof will be in two steps, where the first step is to prove the following auxiliary lemma, and the second step is the concluding corollary immediately following this auxiliary lemma.

**Lemma 3.3** : *Suppose $(X^n)$ is a sequence of processes each of which has zero quadratic variation, and $(Y^n)$ is a sequence of processes each of which has finite quadratic variation and is of the form*

$$Y^n(t) = \sum_{i=0}^{n-1} Y_i^n 1_{(t_i^n, t_{i+1}^n]}(t),$$

*where $Y_i^n$ are a.s. finite random variables and $0 = t_0^n \leq t_1^n \leq ... \leq t_1^n \leq t$. If $(X^n)$ and $(Y^n)$ both converge to $(X_t)$ uniformly on compacts in probability (i.e. both $\sup_{0 \leq s \leq t} |X_s^n - X_s|$ and $\sup_{0 \leq s \leq t} |Y_s^n - X_s|$ converge to zero in probability for each $t \in [0,T]$) then $P\text{-}\lim_{n \to \infty} [Y^n, Y^n] = 0$.*

*Proof:* Since $[X^n, X^n]$ is zero, using the Cauchy-Schwarz inequality we can write $[Y^n, Y^n] = [Y^n - X^n, Y^n - X^n]$. Then,

$$
\begin{aligned}
P\text{-}\lim_{n \to \infty} [Y^n, Y^n]_t &= P\text{-}\lim_{n \to \infty} P\text{-}\lim_{|\Delta| \to 0} \sum \left( (Y_{s_i}^n - X_{s_i}^n) - (Y_{s_{i-1}}^n - X_{s_{i-1}}^n) \right)^2 \quad (3.16) \\
&= P\text{-}\lim_{n \to \infty} P\text{-}\lim_{|\Delta| \to 0} \sum \left[ (Y_{s_i}^n - X_{s_i}^n)^2 - \right. \\
&\qquad \left. 2(Y_{s_i}^n - X_{s_i}^n)(Y_{s_{i-1}}^n - X_{s_{i-1}}^n) + (Y_{s_{i-1}}^n - X_{s_{i-1}}^n)^2 \right],
\end{aligned}
$$

where $\Delta = \{(s_0, s_1), ..., (s_{k-1}, s_k)\}$ is a partition of $[0, t]$.

We see that if we can change the order of the two probability limits we will be able to conclude the result since $|X_t^n - Y_t^n|$ converges in probability to zero uniformly in $t$. We will make use of the fact that the space of real-valued random variables topologized by convergence in probability is a metric space. That is, a sequence of real valued random variables $(Z_n)$ converges in probability to a random variable $Z$ if and only if $d(Z_n, Z) \to 0$ as $n \to \infty$, where the metric $d$ is given by $d(A, B) = E\{|A - B| \wedge 1\}$ for any two random variables $A$ and $B$. Now that we have a metric space, we will make use of the following fact associated with it. *For a doubly-indexed sequence $(x_{mn})$ in a metric space suppose $y_m = \lim_n(x_{mn})$, and $z_n = \lim_m(x_{mn})$ exist for all $m$, $n \in N$, and that the convergence of one of these collections is uniform. Then both the double limit and the iterated limits exist and all three are equal.* We will now show that our case fits into this situation:

$$\text{P-}\lim_{|\Delta| \to 0} \sum \left( (Y_{s_i}^n - X_{s_i}^n) - (Y_{s_{i-1}}^n - X_{s_{i-1}}^n) \right)^2$$

exists $\forall\ n$ and is equal to $[Y^n, Y^n]$, which is well defined due to the discrete nature of $Y^n$. Also,

$$\text{P-}\lim_{n \to \infty} \sum \left( (Y_{s_i}^n - X_{s_i}^n) - (Y_{s_{i-1}}^n - X_{s_{i-1}}^n) \right)^2 = 0,$$

independently of the partition we choose because $|X_t^n - Y_t^n|$ converges to zero in probability uniformly in $t$. Therefore we conclude that the sequences $\Xi_\Delta$ are uniformly convergent, where the sequence $\Xi_\Delta$ is given by

$$\Xi_\Delta = \left( \sum \left( (Y_{s_i}^n - X_{s_i}^n) - (Y_{s_{i-1}}^n - X_{s_{i-1}}^n) \right)^2 \right)_{\Delta, n}.$$

Therefore we can interchange the probability limits in (3.16). Hence

$$\text{P-}\lim_{n \to \infty} [Y^n, Y^n] = 0,$$

which concludes the proof of Lemma 3.3. □

We now can prove the following result.

**Corollary 3.2** : *Suppose $\sigma$ is an a.s. continuous process, and $Z$ is an a.s. continuous process having zero quadratic variation. Then if the process $(\int_0^t \sigma_s dZ_s)$ exists in the sense of (1.1), it has zero quadratic variation.*

*Proof:* Define

$$\sigma^n(t) = \sum \sigma_{t_i^n} 1_{(t_i^n, t_{i+1}^n]},$$

where $0 = t_0^n \leq t_1^n \leq ... \leq t_k^n = t$ and let

$$
\begin{aligned}
T_0^\epsilon &= 0 \\
T_{n+1}^\epsilon &= t \wedge \inf\{s : s > T_n^\epsilon \text{ and } \left|(\sigma \cdot Z)_s - (\sigma \cdot Z)_{T_n^\epsilon}\right| > \epsilon\}
\end{aligned}
\tag{3.17}
$$

be a random partition of [0,t]; further define

$$Y^\epsilon(t) = \sum (\sigma \cdot Z)_{T_n^\epsilon} 1_{(T_n^\epsilon, T_{n+1}^\epsilon]},$$

where we use the notation $(\sigma \cdot Z)_t = \int_0^t \sigma_s dZ_s$. Both $(\sigma^n \cdot Z)$ and $Y_n$ converge to $(\sigma \cdot Z)$ uniformly on compacts in probability. One can easily show that $[\sigma^n \cdot Z, \sigma^n \cdot Z]$ is zero $\forall~n$. Thus by using Lemma 3.3, we have P-$\lim_{n\to\infty}[Y^n, Y^n] = 0$, and this limit equals $[\sigma \cdot Z(\cdot), \sigma \cdot Z(\cdot)]$ by the definition of quadratic variation. Thus, Corollary 3.2 follows. □

This proves Lemma 3.2 since $W$ is a.s. continuous. (Note $W$ is assumed to be a semi-martingale so that $(\int_0^t W_s dZ_s)$ makes sense.) □

We can summarize the above results in the following corollary.

**Corollary 3.3** : *Suppose $\sigma$ is an a.s. continuous adapted semi-martingale, and $Z$ is an a.s. continuous and adapted process having zero quadratic variation. Then $(\int_0^t \sigma_s dZ_s)_{t \in [0,T]}$ exists, has zero quadratic variation, and is a.s. continuous.*

Now we can restate Theorem 2.1 for the particular case of semi-martingale volatility as the following corollary.

**Theorem 3.1** (Arbitrage When the Volatility is a Semi-Martingale)*: Consider the following model for the market:*

$$X_t = \exp(rt)$$

$$dY_t = Y_t \left( \nu dt + \sigma_t dZ_t \right),$$

*where $\sigma$ is an a.s. continuous semi-martingale adapted to the natural filtration of $Z$, which is a process of zero quadratic variation. Then, for each $c > 0$, the portfolio (2.11) is an arbitrage strategy.*

The following corollary relates our proposed model to the existing stochastic volatility models in the literature:

**Corollary 3.4** *: Suppose the dynamics of the stock price are modeled by the following stochastic differential equations:*

$$dY_t = Y_t \left( \nu dt + \phi(W_t) dZ_t \right)$$

$$dW_t = \mu_W(t, W_t) dt + \sigma_W(t, W_t) dB_t,$$

*where $Z$ is a fractional Brownian motion with $H \in (\frac{1}{2}, 1]$, and $B$ is a Brownian motion. (The filtration considered here is either generated by the zero quadratic variation process $Z$ and the Brownian motion, or it is finer.) Assume that $\int_0^T \mu_W(s, W_s) ds < \infty$ and $\int_0^T (\sigma_W(s, W_s))^2 ds < \infty$, and that $\phi \in C^2$. Then, for each $c > 0$, the following portfolio strategy is an arbitrage strategy:*

$$\vartheta_t^0 = cY_0 \left( 1 - \exp \left( 2(\nu - r)t + 2 \int_0^t \phi(W_s) dZ_s \right) \right)$$

$$\vartheta_t^1 = 2c \left( \exp \left( (\nu - r)t + \int_0^t \phi(W_s) dZ_s \right) - 1 \right).$$

*Remark*: The assumptions $\int_0^T \mu_W(s, W_s)ds < \infty$ and $\int_0^T (\sigma_W(s, W_s))^2 ds < \infty$ assure that $W$ is a continuous semi-martingale.

## 3.2 WHEN THE VOLATILITY IS A FUNCTION OF $Z$

We now turn to the situation in which the volatility is a function of the integrator $B^H$ in (1.1). We first give the following lemma of Lin (1995) which states that we can integrate a continuously differentiable function of a continuous process with zero quadratic variation with respect to itself, and moreover that the resulting process is continuous.

**Lemma 3.4** (***Lin (1995)***): *Suppose $Z$ is an a.s. continuous process having zero quadratic variation. Let $\Delta = \{(t_0, t_1), (t_1, t_2), ..., (t_{n-1}, t_n)\}$ be any partition of $[0, t]$. Then for any $C^1$ function $\phi : \mathbb{R} \to [0, \mathrm{T}]$ we have*

$$\int_0^t \phi(Z_s)dZ_s \triangleq \lim_{|\Delta| \to 0} \sum_{(t_{i-1}, t_i)} \phi(Z_{t_{i-1}})(Z_{t_i} - Z_{t_{i-1}}) = \int_0^{Z_t} \phi(x)dx, \qquad (3.18)$$

*where the limit is taken in probability.*

The following lemma shows that the integral in (3.18) has zero quadratic variation. The proof of this lemma works exactly the same way as in Corollary 3.2.

**Lemma 3.5** *Suppose $\phi$ is a continuous function, and $Z$ is an a.s. continuous process of zero quadratic variation. Then $\int_0^t \phi(Z_s)dZ_s$ has zero quadratic variation.*

We summarize the results of this section in the following.

**Theorem 3.2** (ARBITRAGE WHEN THE VOLATILITY IS A FUNCTION OF THE MODULATOR): *Suppose the dynamics of the stock price are modeled by the following differential equation:*

$$dY_t = Y_t (\nu dt + \phi(Z_t)dZ_t),$$

*where $Z$ is a process of zero quadratic variation and $\phi$ is a continuously differentiable function on $\mathbb{R}$. Then, for each $c > 0$, the portfolio of (2.11) is an arbitrage strategy.*

Note that fractional Brownian motion in this theorem can be replaced by any a.s. continuous process having zero quadratic variation, as before.

# 4  CONCLUSION

We have shown that if the financial market is modeled by the modified Black-Scholes equation

$$
\begin{aligned}
X_t &= \exp(rt) &&, \ 0 \leq t \leq T \\
dY_t &= Y_t \left( \nu dt + \sigma_t dZ_t \right) &&, \ 0 \leq t \leq T
\end{aligned}
$$

where $Z$ is either a fractional Brownian motion (fBm) with $H \in (\frac{1}{2}, 1]$ or a time change of it, and $\sigma$ is either an element in the family of adapted continuous semi-martingales, or is a function of the $Z$ itself, then (2.11) is an arbitrage strategy. The arbitrage strategy we provide depends on the Hurst parameter only through the stock price, hence one does not need to estimate the value of $H$ to implement the arbitrage strategy.

The existence of arbitrage strategies for fractional B-S models does not rule out these models as candidates for stock price modeling. One must develop a data oriented approach for modeling the price fluctuations as is suggested by Cutland et al. (1995). Bayraktar et al. (2004) showed that the S&P 500 index possesses long range dependence, by employing an estimator that is robust to seasonalities and volatility persistence and is asymptotically unbiased and efficient. Moreover, the arbitrage strategy that we have constructed in this paper is a continuous trading strategy and thus would not work in a market with transaction costs. Furthermore, for a fractional B-S model with log-normal marginals, Cheridito (2003) has shown shown that arbitrage opportunities do not exist if there is a minimal amount of time $h > 0$ between two consecutive transactions. (Here $h$ can be arbitrarily small.)

Geometric Brownian motion can be justified in a rational expectations equilibrium with highly sophisticated and completely rational agents who instantaneously incorporate all available information into the present price (Kreps (1982) and Bick (1987)). As pointed out by Föllmer and Schweizer (1993), in this approach the beliefs and the preferences of the agents must be specified in a delicate way, and this makes geometric Brownian motion questionable as a robust reference model. Moreover the responses of the agents cannot be instantaneous in practice since information is costly (Grossman and Stiglitz (1980)). Moreover, as is argued by Peters (1994), in a market with participants with different time horizons, it is difficult to interpret what the value of the fair price should be for the market. The fair price for a market participant with a longer time horizon is different for a participant with a short horizon. Therefore an equilibrium on which the participants agree may not make much sense. This also suggests that fractal processes (e.g. fBm) should be considered for modeling stock price fluctuations.

FBm modulated models seem to be more robust as reference models since there are micro structure models justifying the fBm modulated diffusions in the limit. Bayraktar, Horst and Sircar (2003) study the effect of investor inertia on stock price fluctuations with a market microstructure model comprising many small investors who are inactive most of the time. They show that when the price is driven by market imbalance the log price process can be approximated by an integral of a semi-martingale with respect to fBm. Another recent paper that proposes another economic foundation for models based on fBm is Kluppelberg and Kuhn (2002).

The models considered so far in mathematical finance assume that the traders in the market have no price impact. This might be true for the small traders in the market but certainly not true for institutional traders. And in the fundamental theorem of asset pricing (see Delbaen and Schachermayer (1994)) the arbitrageur does not have a price impact even if he has the opportunity of making infinite gain. It is possible to set up a framework where one can study investors with price impacts together with the small investors. The investors with price impact find themselves in a random environment due to the trading noise of the small investors which now enters as a fractional noise, and in accordance with their utilities they control the drift and the volatility coefficients of a stochastic differential equation with fractional Brownian differential. In this setting the observed prices are the Nash-equilibrium prices. Bayraktar and Poor (2002) find the Nash-equilibrium price explicitly in some cases. FBm based models, whose candidacy is justified by empirical evidence, do not fit into the usual framework of mathematical finance; however as we have argued above, it possible to build models that makes it possible to use fBm as as a building block of models for stock prices.

# Acknowledgements

This work was supported by the U.S. Office of Naval Research under Grant No. N00014-03-1-0102.